\documentclass{article}
\usepackage{spconf,amsmath,graphicx}
\usepackage{hyperref, setspace}
\usepackage{multirow, multicol, booktabs}
\usepackage[nolist]{acronym}
\usepackage{cite}
\usepackage{adjustbox}
\usepackage{listings}
\usepackage{xcolor}
\usepackage{amsfonts}
\usepackage{float}

\usepackage{lipsum}
\hypersetup{
    colorlinks=true,
    linkcolor=purple,
    filecolor=magenta,      
    urlcolor=purple,
}

\newcommand{\newpara}[1]{\vspace{6pt}\noindent\textbf{#1}}

\definecolor{codegreen}{rgb}{0,0.6,0}
\definecolor{codegray}{rgb}{0.5,0.5,0.5}
\definecolor{codepurple}{rgb}{0.58,0,0.82}
\definecolor{backcolour}{rgb}{0.95,0.95,0.92}

\makeatletter
\def\bstctlcite{\@ifnextchar[{\@bstctlcite}{\@bstctlcite[@auxout]}}
\def\@bstctlcite[#1]#2{\@bsphack
  \@for\@citeb:=#2\do{%
    \edef\@citeb{\expandafter\@firstofone\@citeb}%
    \if@filesw\immediate\write\csname #1\endcsname{\string\citation{\@citeb}}\fi}%
  \@esphack}
\makeatother 

\usepackage{inconsolata}
\lstdefinestyle{mystyle}{
    backgroundcolor=\color{backcolour},   
    commentstyle=\color{codegreen},
    keywordstyle=\color{magenta},
    numberstyle=\tiny\color{codegray},
    stringstyle=\color{codepurple},
    basicstyle=\ttfamily\footnotesize,
    breakatwhitespace=false,         
    breaklines=true,                 
    captionpos=b,                    
    keepspaces=true,                 
    numbers=left,                    
    numbersep=5pt,                  
    showspaces=false,                
    showstringspaces=false,
    xleftmargin=14pt,
    framexleftmargin=14pt,
    showtabs=false,                  
    tabsize=2
}

\colorlet{punct}{red!60!black}
\definecolor{background}{HTML}{EEEEEE}
\definecolor{delim}{RGB}{20,105,176}
\colorlet{numb}{magenta!60!black}

\lstdefinelanguage{json}{
    basicstyle=\normalfont\ttfamily,
    numbers=left,
    numberstyle=\tiny\color{codegray},
    xleftmargin=14pt,
    framexleftmargin=14pt,
    stepnumber=1,
    numbersep=8pt,
    showstringspaces=false,
    breaklines=true,
    frame=lines,
    backgroundcolor=\color{backcolour},
    literate=
     *{0}{{{\color{numb}0}}}{1}
      {1}{{{\color{numb}1}}}{1}
      {2}{{{\color{numb}2}}}{1}
      {3}{{{\color{numb}3}}}{1}
      {4}{{{\color{numb}4}}}{1}
      {5}{{{\color{numb}5}}}{1}
      {6}{{{\color{numb}6}}}{1}
      {7}{{{\color{numb}7}}}{1}
      {8}{{{\color{numb}8}}}{1}
      {9}{{{\color{numb}9}}}{1}
      {:}{{{\color{punct}{:}}}}{1}
      {,}{{{\color{punct}{,}}}}{1}
      {\{}{{{\color{delim}{\{}}}}{1}
      {\}}{{{\color{delim}{\}}}}}{1}
      {[}{{{\color{delim}{[}}}}{1}
      {]}{{{\color{delim}{]}}}}{1},
}

\newacro{DER}[DER]{diarization error rate}
\newacro{JER}[JER]{jaccard error rate}
\newacro{EV}[EV]{envelope variance}
\newacro{DNN}[DNN]{deep neural network}
\newacro{WER}[WER]{word error rate}
\newacro{SSL}[SSL]{self-supervised learning}
\newacro{SNR}[SNR]{signal-to-noise ratio}
\newacro{SSLR}[SSLR]{self-supervised learning representation}
\newacro{GSS}[GSS]{guided source separation}
\newacro{SOT}[SOT]{serialized output training}
\newacro{SSE}[SSE]{speech separation and enhancement}
\newacro{SA-WER}[SA-WER]{speaker-attributed word error rate}
\newacro{DA-WER}[DA-WER]{diarization-attributed word error rate}
\newacro{CTC}[CTC]{connectionist temporal classification}
\newacro{FA}[FA]{forced alignment}
\newacro{cpWER}[cpWER]{concatenated minimum permutation word error rate}
\newacro{VAD}[VAD]{voice activity detection}
\newacro{AHC}[AHC]{agglomerative hierarchical clustering}

\renewenvironment{thebibliography}[1]{
  \begin{oldthebibliography}{#1}
    \setlength{\itemsep}{0em}
    \setlength{\parskip}{0em}
}
{
  \end{oldthebibliography}
}
\title{One model to rule them all ? Towards End-to-End joint speaker diarization and speech recognition}
\name{Samuele Cornell$^{1,2}$\thanks{S. Cornell was partially supported by the ``Miracle'' project POR MARCHE FESR 2014-2020.}, Jee-weon Jung$^{2}$, Shinji Watanabe$^{2}$,  Stefano Squartini$^{1}$}
\address{ $^1$Università Politecnica delle Marche, Italy
  $^2$Carnegie Mellon University, USA \\ }
\begin{document}
\ninept
\maketitle
\begin{abstract}
This paper presents a novel framework for joint speaker diarization (SD) and automatic speech recognition (ASR), named SLIDAR (sliding-window diarization-augmented recognition). 
SLIDAR can process arbitrary length inputs and can handle any number of speakers, effectively solving ``who spoke what, when'' concurrently. 
SLIDAR leverages a sliding window approach and consists of an end-to-end diarization-augmented speech transcription (E2E DAST) model which provides, locally, for each window: transcripts, diarization and speaker embeddings.
The E2E DAST model is based on an encoder-decoder architecture and leverages recent techniques such as serialized output training and ``Whisper-style" prompting. 
The local outputs are then combined to get the final SD+ASR result by clustering the speaker embeddings to get global speaker identities.
Experiments performed on monaural recordings from the AMI corpus confirm the effectiveness of the method in both close-talk and far-field speech scenarios. 
%It allows for efficient and effective speaker-attributed and segmented transcription of recordings with an arbitrary number of participants, 
% It incorporates several novel elements, including a SOT objective with joint speaker diarization, an efficient inference scheme for multi-talker speech based on sliding windows, and additional multi-task learning objectives enabled by decoder prompting. 
% 112 words
%In detail, FIXME % add detailed results 
% note: 100 to 150 words
\end{abstract}

\section{Introduction}
\label{sec:intro}

Speaker diarization (SD) and automatic speech recognition (ASR) are both essential tasks in many applications, such as meeting transcription, speech analytics, automatic captioning, and speech-enabled virtual assistants to name a few~\cite{garofolo2002nist, park2022review, prabhavalkar2023end,
barker2018fifth, watanabe2020chime}. 
These tasks, being mutually complementary, have often been performed together (SD+ASR). %SD, for example, was historically used for speaker-adaptation of ASR systems~\cite{park2022review, serafini2023experimental} in order to boost recognition performance. 
%but recent works are also exploring using ASR outputs for improving SD~\cite{}. not needed here. % and in particular ones involving multiple speakers,
This integration stems from the fact that, in actual use-cases, it is often necessary to obtain a ``who spoke what, when'' diarization-augmented speech transcription (DAST)~\cite{garofolo2002nist, fiscus2007rich, cornell2023chime, yu2022m2met}: i.e. a  ``rich'' speech transcription with segmentation and speaker-attribution at the utterance or word-level. %Research efforts towards achieving such capabilities span more than two decades. %Notable benchmarks and evaluation criteria, such as NIST Rich Transcript Evaluation~\cite{garofolo2002nist} as well as challenges and datasets like recent CH\-iME editions~\cite{barker2018fifth, watanabe2020chime, cornell2023chime}, Alimeeting~\cite{yu2022m2met}, AMI~\cite{carletta2005ami} to name a few, had a significant role in this direction. 

Such DAST output for long-form audio recordings is generally achieved by using a complex SD+ASR pipeline which comprises of, generally: SD, speech separation/enhancement (SSE) and ASR components~\cite{medennikov2020stc, du2020ustc, raj2021integration, yu2022summary, cornell2023chime, boeddeker2023ts}. 
However, such an approach has an inherent drawback: it is difficult to optimize jointly all the modules. This leads to sub-optimal overall performance and the need for lots of hyper-parameter tuning including, but not limited to, SD output smoothing~\cite{medennikov2020stc, cornell2023chime, boeddeker2023ts}.
This is also especially true when dealing with monaural recordings (which are also more commonly available), as it makes separation and diarization way more challenging.
 %only monaural recordings, which are arguably more available, are available as separation and diarization become much more challenging. 
%: e.g. separation can't leverage multi-channel constrained solutions and becomes susceptible to introducing mismatches for the ASR system~\cite{}. %unless the ASR model is re-trained or joint SSE+ASR fine-tuning is performed~\cite{}. 
%In such scenarios, separation becomes more challenging and is susceptible to introducing mismatches for the ASR system~\cite{} and joint SSE+ASR fine-tuning or ASR re-training are necessary~\cite{chang2022end}.
%but then the separation output is not guaranteed to be suitable for diarization too. 
%multi-talker long-form %the drawback that it is difficult to optimize jointly 

Recent works have aimed to reduce the components within the SD+ASR pipeline, for example, by trying to combine closely related tasks such as SD and speech separation~\cite{maiti2023eend}.
Indeed, some works~\cite{chen2020continuous, yoshioka2022vararray, morrone2023low} try to replace SD altogether with SSE via continuous source separation (CSS)~\cite{chen2020continuous}. However, such approaches are not suitable for long-form audio. CSS will lose the speaker tracking when two or more speakers are not active within a CSS window overlap and additional SD is thus necessary~\cite{raj2021integration}. As such e.g. \cite{yoshioka2022vararray, kanda2022vararray} were only evaluated on utterance groups separately or \cite{morrone2023low} focused only on 2-speakers scenarios and relatively short recordings. The same problem also applies to all ASR-based systems that rely only on permutation invariant training (PIT)~\cite{kolbaek2017multitalker}, SOT~\cite{kanda2020serialized} or other local-only strategies to perform multi-speaker ASR~\cite{chang2020end, kanda2020serialized, lu2021streaming, raj2022continuous,
Kanda2022StreamingMA}. 
% sam: bibliography is heavily biased towards microsoft work

Target-speaker approaches do not suffer from such a problem but require a prior SD output. As such, target-speaker SSE methods~\cite{boeddeker2018front, delcroix2018single, boeddeker2023ts}, while being effective, do not really simplify the pipeline. 
Instead, ASR-based target-speaker methods~\cite{kanda2020joint, kanda2021investigation, kanda2022streaming, moriya2022streaming, huang2023adapting} can reduce, in principle, the components only to SD+ASR with the ASR doing implicitly speech separation. For example, Transcribe-to-diarize (T2D)~\cite{kanda2022transcribe} simplifies considerably the pipeline as it needs only \ac{VAD}, clustering-based SD and an E2E speaker-attributed ASR model~\cite{kanda2020joint}. This latter utilizes pre-estimated speaker embeddings obtained from the clustering-based SD step to generate a DAST with consistent speaker-ids across all the recordings. Contrary to aforementioned works on target-speaker ASR~\cite{kanda2020joint, kanda2021investigation, kanda2022streaming, huang2023adapting} here the ASR model does not only do speaker-attribution but also outputs a refined diarization output.

\begin{figure*}[t]
\centering
\includegraphics[width=0.59\textwidth]{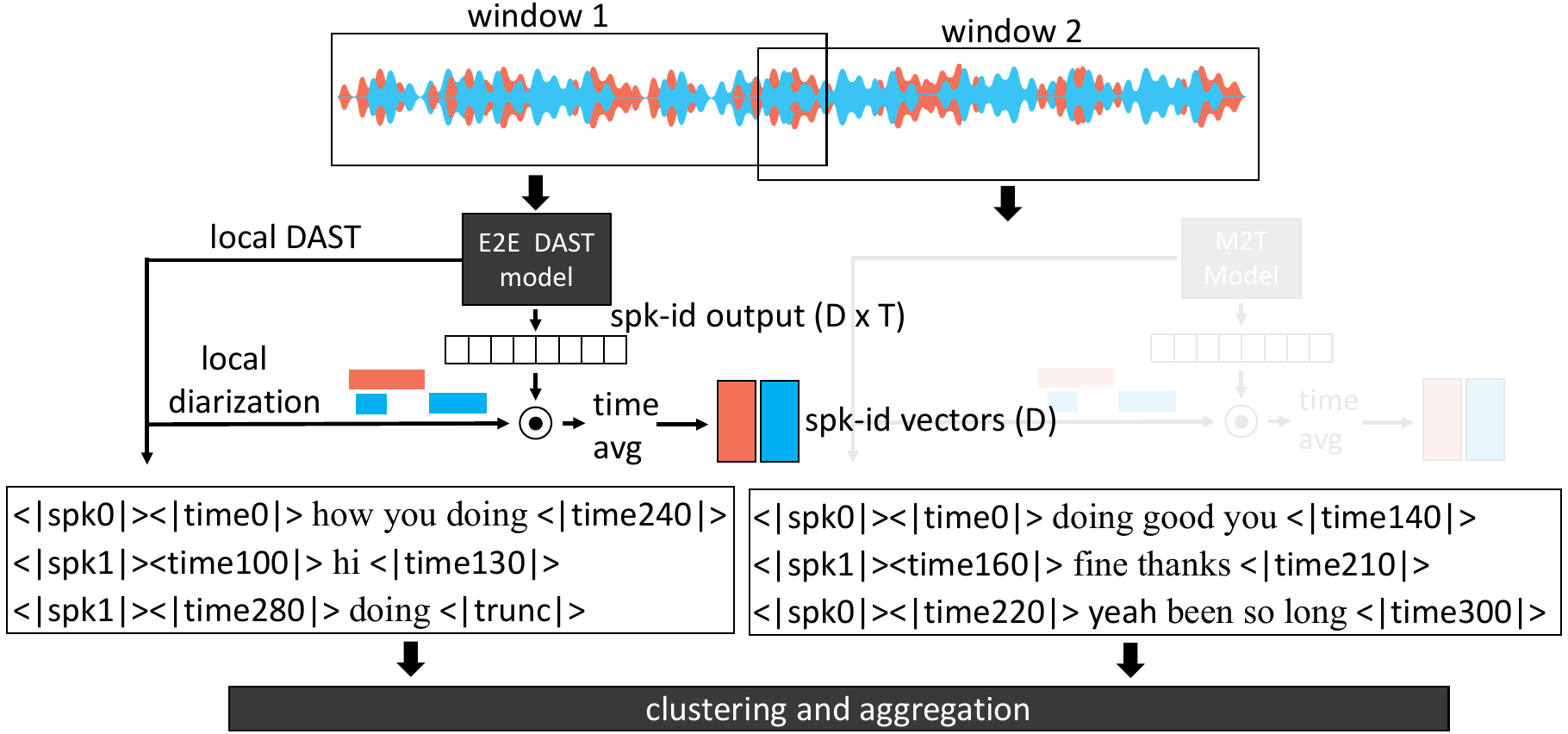}
\vspace{-0.4cm}
\caption{Proposed SLIDAR framework on a two speakers toy example.}
\label{fig:sassr_framework}
\vspace{-0.4cm}
\end{figure*}

%Instead of leveraging several sub-components, an intriguing approach is to try to devise a single model able to seamlessly segment, diarize and transcribe a long-form recording with possibly multiple speakers. 
In this work, we propose a relatively simple yet effective framework that solves the problem of ``who spoke what, when'' simultaneously. We call this approach SLIDAR as it jointly performs speaker-attribution, speech segmentation, and recognition in a ``quasi" end-to-end (E2E) manner, as it uses a sliding window and clustering mechanism to keep the computational cost linear with the length of the recording. 
%a clustering mechanism is employed. 
SLIDAR first adopts an E2E DAST attention-based encoder/decoder (AED)~\cite{prabhavalkar2023end} model to derive local transcriptions plus SD as well as speaker embeddings.  
Following, all results from each window are combined and final global speaker labels are assigned by clustering all the local speaker embeddings. 
Unlike T2D, which is the most closely related work to ours, SLIDAR does not require any prior \ac{VAD} or clustering-based diarization output. 
%To the our best knowledge this is the first time an almost e2e system 
%Then a sliding window mechanism is applied on top of local results with a clustering for global speaker labels. 
%Note that local speaker labels are only used during training. Global speaker labels are derived using the embeddings.
%on a long-form input audio recording and integrates fully E2E ``local" DAST with a final clustering step which assigns speaker labels consistent across all recording. 
This framework relies on multiple novel contributions: (i) E2E joint SD+ASR modeling with serialized output training (SOT)~\cite{kanda2020serialized}, enabled by the addition of several special tokens; (ii) a sliding-window based inference mechanism suitable for multi-talker recordings; (iii) joint speaker embeddings estimation via the SOT SD+ASR output; (iv) new ``Whisper-style" multi-task learning tasks for multi-talker long-form recordings.

Experiments and ablation studies (Sec.~\ref{sec:exp_valid}) on real-world monaural long-form multi-talker recordings i.e. AMI headset-mix and far-field microphone signals indicate promising potential for this method, with comparable or better performance with respect to state-of-the-art methods such as T2D.

\section{SLIDAR framework}
\label{sec:format}

Figure~\ref{fig:sassr_framework} illustrates our proposed framework, SLIDAR.
SLIDAR employs a sliding window mechanism on an input long-form recording, where each window is processed independently by a local E2E DAST model.
Such a mechanism ensures that our approach can be applied to an arbitrary length input while keeping the memory requirements constant, enabling the utilization of large pre-trained models (these usually employ self-attention, e.g. WavLM~\cite{chen2022wavlm}). 
However, this also requires to have a technique that keeps track of the speaker identities globally. In this framework, the E2E DAST model is tasked with estimating, for each window, a ``local" DAST output as well as speaker embeddings of speakers. 
The ``global" transcription is derived by aggregating local transcriptions; global speaker labels are computed by clustering all speaker embeddings.

%As such, the proposed framework is fully E2E except for this latter clustering operation.
%lev
%with the length of the whole recording, making it possible to leverage large pre-trained models 
%Such strategy  quite effective for diarization even for short windows (e.g. 5\,s) 

\subsection{Local E2E DAST Model}\label{ssec:local_dast}

The local E2E DAST model is based on AED and adopts a SOT strategy. It predicts, for each utterance in the current input, special tokens for \textit{onset} and \textit{offset} timestamps, the words and a \textit{speaker tag} token for the speaker to which the utterance belongs: e.g. \texttt{<|spk0|> <|time10|> hello world <|time200|>}, where  \texttt{<|time10|>} and \texttt{<|time200|>} are respectively the \textit{onset} and \textit{offset} timestamp tokens of the utterance. 

Except for these changes, described in detail below, the model is trained following the standard SOT~\cite{kanda2020serialized} framework using teacher forcing at the attention-based decoder output $y_i \in [y_{1},..,y_{M}]$:

\begin{equation}
    \mathcal{L}^{\text{SOT}}=
\sum_{i=1}^{M}{\rm CE}(y_i, l^{\Psi}_{i}).
\end{equation}

Where $M$ is the target length. $\Psi$ defines the particular permutation of the $l_i \in [l_{1},..,l_{M}]$ target sequence, i.e. the target tokens sequence permutation for which each utterance is ordered by its starting time, in a first-in first-out (FIFO) manner (tokens from utterances in Fig.~\ref{fig:sassr_framework} plus an end-of-sequence \texttt{<|eos|>} token). CE denotes the cross-entropy loss function between output distribution and target labels, again as in~\cite{kanda2020serialized}.
% The timestamp symbols are defined as in Whisper, i.e. 

\newpara{Timestamp Tokens.} We use a fixed set of special tokens that represents timestamps, identical to Whisper~\cite{whisper}. 
The amount of these timestamp tokens depends on their resolution (e.g. 100\,ms) as well as the maximum window length we want the model to support in inference.
A \texttt{<|nospeech|>} special token is used when the current window does not contain any speech.

\newpara{Speaker Tag Tokens.} Our approach can be considered an extension of \cite{kanda2020serialized, kanda2022transcribe, Kanda2022StreamingMA}. Instead of adopting a speaker change symbol \texttt{<|sc|>}, we use $N$ different speaker tag tokens: \texttt{<|spk0|>, <|spk1|>},$\hdots,$\texttt{<|spkN|>} with $N$ chosen a-priori as the maximum number of speakers we expect in the local window. 
Unlike earlier studies that rely on a speaker inventory~\cite{Kanda2022StreamingMA} or clustering-based diarization output~\cite{kanda2022transcribe}, here, even at the local level, we have an arbitrary mapping problem between reference speaker names and the estimated speaker tags. 
To overcome this ambiguity, we simply enforce for the speaker tags a FIFO naming convention: within each window, the first speaker to appear is labeled as \texttt{<|spk0|>}, the second as \texttt{<|spk1|>} and so on (see Fig.~\ref{fig:sassr_framework}). Note that such speaker tag consistency is only enforced within the current window.

%In other words, the model is tasked to implicitly identify for the relative speaker in each window. 
%Speaker diarization techniques such as EEND-VC~\cite{kinoshita2021integrating} or the Pyannote diarization pipeline~\cite{bredin2020pyannote} adopt this same local vs. global divide and conquer strategy which allows to handle an arbitrary number of total speakers.  %The idea is closely related to EEND-VC~\cite{kinoshita2021integrating}: we produce a local, permutation invar
%A fundamental problem 

\newpara{Truncation Token.} Since the model is applied (both in inference and training) on finite audio chunks from e.g. a meeting scenario, it can happen that some utterances are truncated from such windowing approach. %This is a fundamental issue that must be resolved if our goal is fully E2E DAST at the local level, without any separate VAD component. 
To handle such cases, we introduce an additional special truncation token (\texttt{<|trunc|>}) (see Fig.~\ref{fig:sassr_framework}) which is used in place of timestamps when an utterance is truncated: e.g. if truncated at the beginning it replaces the onset timestamp, if truncated at its end, the offset timestamp. 
Such \texttt{<|trunc|>} token is key for fully local E2E DAST as it allows the model also to take over the VAD role.
This allows training the model on randomly sampled windows but also requires world-level segmentation in order to split utterances that are truncated. 
Such segmentation can be obtained in a scalable way only via \ac{FA}. Yet FA can be unreliable in the presence of overlapped speech if per-speaker close-talk microphones are not available (e.g. web videos etc.). Thus, here, we also explore a different strategy, where, as in~\cite{von2023meeteval}, we estimate word boundaries from utterance boundaries by counting the characters in each word and assuming these latter isochronal. Our experiments (see Tab.~\ref{tab:ablation_ihm}) show that this form of ``weak supervision" can work reasonably well. %After all, with sufficiently long windows, only a minority of utterances are truncated during training compared to the total seen by the model.

%This, in some scenarios (e.g. large-scale weak supervision training) can be a significant advantage over other methods that rely on word-level segmentation~\cite{kanda2022transcribe}.
% advantage over SOT is that the no decoding on utterance groups 
% advantage over tSOT is that we do not need 

%This is a significant advantage of uSOT over tSOT. 

%is the fact that we demonstrate that it does not strictly need word-level segmentation (see Table.~\ref{tab:ablation_ihm}) during training. 
%which could be unreliable or expensive to obtain in many scenarios. 

%On the other hand, if word-level segmentation is needed, a modified tSOT (with the \textit{speaker} tag symbols etc.) can be employed or the strategy adopted here could be modified by having the model outputting, for each word, a speaker tag together with \textit{onset} and \textit{offset} timestamps.

\subsection{Windowing and Decoding}\label{ssec:inference_algo}
As mentioned, the inference phase employs a sliding window mechanism on the whole recording. 
As such, we have to take into account the fact that the window could truncate the final utterance of multiple speakers at the end-point of the window. I.e., for some speakers, in their last utterance, instead of an \textit{offset} token, a \texttt{<|trunc|>} is encountered.
When this occurs, we enforce the start of the next window at the rightmost silence segment (no speaker active) from the leftmost utterance that was truncated in the previous decoding window (e.g. for two speakers, see Fig.~\ref{fig:sassr_framework}). This assures that no truncation at utterances onset is encountered during inference. %will then be encountered in the next window.  
%We found this simple heuristic to be effective as long as the window is made long enough.
Instead, when no utterances are truncated at the offset, we simply start the next window at the end of the current one (no overlap).  %decoding

Within each window, during decoding and at each autoregressive iteration, we exclude inadmissible hypotheses from the beam search stack. This is crucial for achieving satisfactory performance. 
For example, we enforce rules such as: (i) timestamp tokens should be monotonically increasing for each speaker (ii) after a timestamp token, the next one can be either a speaker token or a subword unit token; and so on, so that all hypotheses are well formed.

\subsection{Speaker Vectors Estimation and Clustering}\label{ssec:spk_id_vect}

Together with the local E2E DAST output, the model is also tasked to estimate speaker embeddings for each window, to be used to assign global speaker labels and merge all predictions.  
In detail, (see Fig~\ref{fig:sassr_framework}), we require the local E2E DAST model encoder also to output frame-level speaker discriminative features via another output head $E(\cdot)$: 
\begin{equation}
    \mathbf{h}_i = E(\mathbf{x}_j)
\end{equation}
Where $\mathbf{x}_{j} \in \mathbb{R}^{D_{f} \times K}$ is the input feature vectors sequence to the encoder with length $K$ and size $D_{f}$ each: $\mathbf{x}_{j} = [\mathbf{x}_1, \hdots, \mathbf{x}_K]$. While
$\mathbf{h}_{i} \in \mathbb{R}^{D_{h} \times M}\,,\, \mathbf{h}_{i} = [\mathbf{h}_1, \hdots, \mathbf{h}_M]$ is the corresponding output sequence with length $M$ and size $D_{f}$ each.  
We use the local DAST diarization information estimated by the decoder via SOT to compute the local speaker embedding vectors $\mathbf{e}_{s} \in \mathbb{R}^{D_{e} \times N}\,,\, \mathbf{e}_{s} = [\mathbf{e}_1, \hdots, \mathbf{e}_N]$ at each window using time-averaging:

\begin{equation}\label{eq:spkid_vect}
    \hat{\mathbf{e}}_{s} = \frac{1}{n(\Theta_{s})}\sum_{i \in \Theta_{s}}  \mathbf{h}_{i},\,\,\,s \in {1, \hdots, N^{\text{local}}}
\end{equation}
where $\Theta_{s} \subseteq \{x \in \mathbb{N} \mid 1 <= x <= M \}$ is the set of frames for which only the $s$-th local speaker is active (thus without considering overlapped speech parts), $n(\Theta_{s})$ is the cardinality of the set and $N^{\text{local}} \leq N$ is the number of local speakers in the current window (see Sec.~\ref{ssec:local_dast}). 
Note that this is equivalent to applying a binary mask and it is a key difference from~\cite{kinoshita2021integrating} where instead a sigmoid mask is employed. 
We train such speaker embedding $E(\cdot)$ encoder head jointly with the local DAST objective $\mathcal{L}^{\text{SOT}}$ by using the $\mathcal{L}^{\text{speaker}}$ speaker embedding loss from EEND-VC~\cite{kinoshita2021integrating, kinoshita2021advances}:
\begin{eqnarray}
\mathcal{L}^{\textrm{speaker}} &=&  \frac{1}{N^{\textrm{local}}} \sum_{s=1}^{N^{\textrm{local}}} l^{\textrm{speaker}} \left({\sigma}_{s}, \hat{\mathbf{e}}_{s} \right). \label{eq:speaker_loss_1} 
\end{eqnarray}
Where $\mathbf{\sigma}_s$ is the $s$-th target speaker identity index over all speakers in the training set and $l^{\textrm{speaker}}$ is defined exactly as in~\cite{kinoshita2021integrating}; an embedding dictionary for all training speakers is learned jointly during training. Note that, in training, we use oracle diarization for time-averaging in Eq.~\ref{eq:spkid_vect} instead of the estimated DAST output, as this latter would require decoding. Also, compared to~\cite{kinoshita2021integrating}, here we have no permutation ambiguity since the local speaker order is FIFO-based. 
%as this latter would require to decode  

Since inference is performed on sliding windows that can overlap, we use the overlapped part information, again as in EEND-VC, to impose cannot-link constraints in the clustering phase. We chose in this work constrained \ac{AHC}~\cite{kinoshita2021advances} and impose constraints only when the overlap is greater than 5\,s.
%We demand to future work the possibility to ensemble also the speakers transcriptions hypotheses in the overlapping portions to improve performance. 

%We extract here speaker embeddings only from non-overlapping regions as in Pyannote SD pipeline.
%Underlying assumption is that if the window is sufficiently long, due to the sparsity of most conversations, we can  obtain reliable speaker embeddings\footnote{The identical assumption is made in Pyannote~\cite{}}.
% The local E2E DAST output diarization information is used to derive speaker discriminative embeddings in a similar manner as employed in the Pyannote diarization pipeline~\cite{}.
% The idea is that if the window is sufficiently long, due to the sparsity of most conversations, it will be possible to obtain reliable speaker vectors by extracting them only from non-overlapped speech regions, as done e.g. in the Pyannote~\cite{} diarization pipeline.
%This has been demonstrated by pyannote papers. 

%We instead use the overlapped part information only in the clustering step as in EEND-VC~\cite{kinoshita2021integrating} to impose constraints in the clustering algorithm. 

\subsection{Multi-Task Training}\label{ssec:multi_task}
Since the proposed framework is fundamentally built upon AED ASR (it uses SOT), we can use the same ASR multi-task learning framework introduced in Whisper~\cite{whisper}.
In particular, here we experiment with two additional tasks/modes on top of the local DAST prediction: (i) \texttt{<|OD|>}, where the model is prompted with oracle diarization information for the current window and (ii) a \texttt{<|vanillaASR|>} task, to allow the model to be trained on non-conversational ASR corpora (which are arguably larger) and also cover single-speaker pre-segmented inputs. On top of these two, we also experiment with Whisper previous decoding window output conditioning (\texttt{<|prev|>}) for the current prediction. 
%These could be further expanded in future to include also diarization-only and word-level vs. utterance-level DAST, with the goal of maximizing the amount and diversity of training data that can be leveraged by the model. 

\section{Experimental Setup}
\label{sec:exp_setup}

\subsection{Dataset: AMI Corpus}\label{ssec:datasets}
The AMI Corpus~\cite{carletta2005ami} consists of over 100~h real-world meetings between 3 to 5 participants recorded by a variety of devices, including distant microphone arrays and close-talk lapel and headset microphones. It is partitioned~\cite{landini2022bayesian} into \texttt{train} (80.2\,h), \texttt{dev} (9.7\,h),
and \texttt{test} (9.1\,h) sets. % with a total of X,Y and Z speakers respectively. 
To compare with \cite{kanda2022transcribe}, we experiment with two conditions: single distant microphone (SDM) and headset-mix (IHM-MIX) recordings. 
Importantly, AMI offers reliable word-level segmentation annotation obtained via \ac{FA} on the worn headset mics, allowing us to compare our SLIDAR strategy with and without the utterance truncation approximation explained in Sec.~\ref{ssec:local_dast}. %As said, this is more an exception than the rule, as solid \ac{FA} is difficult without close-talk microphones. 

%Ground truth speaker activity was obtained by human annotators from close-talk speaker-worn microphones while distant speech was recorded by two 8-microphone circular arrays, each with a 10\,cm diameter: one placed at the end and another at the centre of the meeting table used by the participants.
%\subsubsection{Mixer 6 Speech}

\subsection{Model Architecture}\label{ssec:model_arc}
Our framework can employ any AED ASR model with minor modifications (e.g. including Whisper). In this work, we adopt the WavLM-supported Transformer AED model introduced in IRIS~\cite{chang2022end} but without any \ac{CTC}~\cite{graves2006connectionist} loss on the encoder. Instead, the encoder here has a separate linear layer to output $256$ dimensional speaker discriminative features ($\mathbf{h}_i$, see Sec.~\ref{ssec:spk_id_vect}). 
Compared to~\cite{chang2022end} we use $2$ encoder and $6$ decoder layers. Each an internal representation of size $512$, $4$ attention heads, $0.2$ dropout probability and $1024$ as the hidden dimension of the feed-forward layer.
We use SentencePiece~\cite{kudo2018sentencepiece} tokenization and set the vocabulary size to $500$ non-special tokens. 
The timestamps special tokens have a time resolution of $100$\,ms and a max duration of $20$\,s. The number of speaker tag tokens is $5$. 

%Thanks to the sliding window inference approach, the model can transcribe a $1$\,h recording in approximately $5$\,minutes using a single A100 40GB NVIDIA GPU. 
%but the $\mathcal{L}^{speaker}$ speaker embedding loss as said in Sec.~\ref{ssec:spk_id_vect}.
%as it has shown to be particularly effective~\cite{masuyama2022end, cornell2023chime}.  %and found very effective on the CHiME-7 DASR challenge~\cite{}.  
%The only differences is that here we do not employ a and, since we want the encoder to encode the 

%Due to the sliding window approach, it can achieve a real-time factor of 10 on a 
%Say how much fast is inference. 

\begin{table*}[h!]
%\ra{1.1}
% \setcounter{table}{4}
  \caption{DER (\%) and cpWER (\%) on the AMI corpus \texttt{dev} and \texttt{eval} sets. 
  DER is computed without any collar as in~\cite{kanda2022transcribe}; \\ For SLIDAR we also report the results obtained with \textit{oracle diarization} prompting (\texttt{<|OD|>} mode) during inference (see Sec.~\ref{ssec:multi_task}). \\ SC: speaker confusion (\%) , MS: missed detection (\%) , FA: false alarm (\%).}
  \label{tab:ami_summary}
  \vspace{0.5mm}
  \centering
{\footnotesize
\begin{tabular}{@{}llllllll@{}}
    \toprule
Mic & System & Configuration  &  \multicolumn{2}{c}{dev} && \multicolumn{2}{c}{eval} \\ \cmidrule{4-5} \cmidrule{7-8} 
% &       &               &  SE\hspace{1.6mm} / Miss\hspace{0.8mm} / FA \hspace{1.2mm}/ DER & cpWER && SE / Miss / FA / DER & cpWER  \\ \midrule
&       &               &  \multicolumn{1}{c}{SC / MS / FA / DER} & cpWER && \multicolumn{1}{c}{SC / MS / FA / DER} & cpWER  \\ \cmidrule{1-8}
 & VBx \cite{landini2022bayesian} & oracle VAD & 2.88 / 13.45 / 0.00 / 16.33  & $\:\:\:$- && 4.43 / 14.56 / 0.00 / 18.99 & $\:\:\:$- \\ 
%IHM-MIX & WavLM-ASR & oracle separation &  & - &&  & - \\ \midrule
%IHM-MIX & SC~\cite{kanda2022transcribe} &  & 3.37 / 14.89 / 9.67 / 27.93 & 23.1 &&  3.45 / 16.34 / 9.53 / 29.32  & 23.4\\
% keep-full
% IHM-MIX & SC $\rightarrow$ E2E SA-ASR & estimated & 3.10 / 11.61 / 9.83 / 24.54 & 16.1 && 2.56 / 13.92 / 8.39 / 24.86 & 16.6 \\ 
% IHM-MIX & SC $\rightarrow$ E2E SA-ASR & oracle$^\dagger$ & 3.10 / 11.61 / 0.00 / 14.71  & - && 2.56 / 13.92 / 0.00 / 16.48 & - \\ \midrule
% vad
 & 
T2D~\cite{kanda2022transcribe}
%SC $\rightarrow$ E2E SA-ASR 
&         &  3.05 / 11.46 / 9.00 / 23.51 & 15.9 && 2.47 / 14.24 / 7.72 / 24.43 & 16.4 \\ 
% IHM-MIX & SC $\rightarrow$ E2E SA-ASR & oracle$^\dagger$ &  3.05 / 11.46 / 0.00 / 14.50 & - && 2.47 / 14.24 / 0.00 / 16.71 & - \\ \midrule
IHM-MIX & 
T2D~\cite{kanda2022transcribe}
%SC $\rightarrow$ E2E SA-ASR 
& oracle VAD &  2.83 / $\:$ 9.69 / 3.46 / 15.98 & 16.3 && 1.78 / 11.71 / 3.10 / 16.58 & 15.1 \\ 

 & 
SLIDAR
%SC $\rightarrow$ E2E SA-ASR 
&   & 4.02 / 7.39 / 13.81 / 25.22  & 14.2 &  & 3.46 / 5.74 / 17.90 / 27.10  & 15.6 \\

 & 
SLIDAR
%SC $\rightarrow$ E2E SA-ASR 
&  oracle diarization  & \hfil - & 10.8 && \hfil - & 11.5 \\ 
% &  SLIDAR
%SC $\rightarrow$ E2E SA-ASR 
%&  two-pass decoding  &  & 13.7   && & \\ 
\cmidrule{1-8}
%SDM & SC~\cite{kanda2022transcribe} &   &  3.50 / 21.93 / 4.54 / 29.97 & 28.6 && 3.69 / 24.84 / 4.14 / 32.68 & 30.3\\
% keep-full
%     SDM & SC $\rightarrow$ E2E SA-ASR & estimated & 3.95 / 12.97 / 9.38 / 26.29 & 22.0 && 3.50 / 15.21 / 8.28 / 26.98 & 23.9 \\ 
% SDM & SC $\rightarrow$ E2E SA-ASR & oracle  & 3.95 / 12.97 / 0.00 / 16.92 & - && 3.50 / 15.21 / 0.00 / 18.71  & - \\ \bottomrule 

 & 
    VBx~\cite{landini2022bayesian}$^\ddagger$  &  & \hfil - & \,\,\,\,- && 4.83 / 18.15 / 3.24 / 26.22   & \,\,\,\,- \\ 
&  SC + OVL~\cite{raj2021multi}$^\ddagger$
   % SC $\rightarrow$ E2E SA-ASR 
    &  & \hfil - & \,\,\,\,- &&  6.67 / $\:$9.63 / $\:$7.39 / 23.69  & \,\,\,\,- \\ 
% vad
     & 
    T2D 
   % SC $\rightarrow$ E2E SA-ASR 
    &  & 3.48 / 15.93 / 7.17 / 26.58 & 22.6 && 2.86 / 19.20 / 6.07 / 28.12 & 24.9 \\ 
% SDM & SC $\rightarrow$ E2E SA-ASR & oracle  & 3.48 / 15.93 / 0.00 / 19.41  & - && 2.85/ 19.20 / 0.00 / 22.05  & - \\ \bottomrule
SDM & 
T2D
%SC $\rightarrow$ E2E SA-ASR 
& oracle VAD  & 3.38 / 10.62 / 3.28 / 17.27  & 21.5 && 2.69 / 12.82 / 3.04 / 18.54  & 22.2 \\ 
 & 
SLIDAR
%SC $\rightarrow$ E2E SA-ASR 
&   & 4.92 / 14.53 / 9.83 / 29.28   & 21.8 && 3.79 / 19.73 /  8.01 / 31.52  & 24.5 \\

 & 
SLIDAR
%SC $\rightarrow$ E2E SA-ASR 
& oracle diarization  & \hfil - & 19.4 && \hfil - & 21.1 \\ 
% &  SLIDAR
%SC $\rightarrow$ E2E SA-ASR 
%&  two-pass decoding  &  &  && &  \\ 
\bottomrule

  \end{tabular}
  }\\
{\footnotesize \hspace{-10cm} $^\ddagger$ obtained from \href{https://github.com/desh2608/diarizer/tree/master}{{github.com/desh2608/diarizer}}}.\\
  %\vspace{-6.5mm}
  \vspace{-5mm}
\end{table*}

\subsection{Training and Inference Details}\label{ssec:training}
%We train the model with variable windows sizes.
We train the model by using the full AMI training corpus (all microphones, including all arrays), CH\-iME-6~\cite{watanabe2020chime} (all far-field microphones only) and 5k\,h of simulated $3$ to $5$ speakers meetings obtained using LibriSpeech~\cite{panayotov2015librispeech}, SINS~\cite{dekkers2017sins}, and Pyroomacoustics~\cite{scheibler2018pyroomacoustics} by using NVIDIA NeMo synthetic data simulator~\cite{nemo_2019}. For these, we randomly sampled room size between $20-50$ $m^2$ and T60 reverberation time between $0.3-0.7$ s. The microphone position is randomly sampled inside the room and then kept fixed while speaker positions are sampled at each new utterance. %In training we use as input anechoic noisy, anechoic clean, reverberant noisy and reverberant clean signals. 
%For these latter we produce far-field noisy, far-field clean, close-talk noisy and close-talk clean examples to the model. actual data is then 4 times 5k hours 

During training, we randomly sample $20$\,s windows from each recording and use with, respectively $0.5$ and $0.1$ probability, \texttt{<|prev|>} and \texttt{<|OD|>} multi-task prompt conditioning modes. For truncation, we use the approximation outlined in Sec.~\ref{ssec:local_dast} from AMI, CHiME-6, and LibriSpeech manual segmentation unless stated otherwise. 
When multi-task training with \texttt{<|vanillaASR|>}, we use additional data from the full Mixer 6 Speech~\cite{brandschain2010mixer} training set as well as pre-segmented utterances from all the previously mentioned corpora. 

We use a dynamic batch size with a total of $900$\,s of input audio examples, Adam optimizer~\cite{kingma2014adam}, a learning rate of $1e^{-3}$ and $8$ GPUs. 
The model is trained for $200$k steps, keeping WavLM frozen. 
After this, we fine-tune the AED model and WavLM together with a learning rate of $1e^{-6}$ for additional $20$k steps, using gradient checkpointing to not incur in out-of-memory issues. We use adaptive SpecAugment~\cite{park2020specaugment} on the WavLM extracted features as in~\cite{chang2022end}, clip gradients $l_2$ norm to $5$, and adopt a linear learning rate schedule with $20k$ warm-up steps and then decay. 
In inference, we use $20$\,s sliding window (hop size variable see Sec.~\ref{ssec:inference_algo}) and beam-size $10$ for decoding.

\section{Experimental Results}
\label{sec:exp_valid}
We evaluate our proposed framework in terms of \ac{DER} and \ac{cpWER}~\cite{watanabe2020chime} to assess both SD and ASR performance. We adopt the same evaluation protocol and AMI ground truth segmentation as used in~\cite{landini2022bayesian, kanda2022transcribe}.  %and in particular T2D~\cite{kanda2022transcribe}, as well as previous works that focused only on diarization. 

%In Figure~\ref{fig:perf_vs_window} we investigate how inference window length affects performance on AMI-IHM. Recall that the model was trained (see Sec.~\ref{ssec:training}) with variable length windows between (20\,s and 2\,mins) and thus can support variable length inference windows. 

In Table~\ref{tab:ami_summary}, we compare, on AMI IHM-MIX and SDM recordings, SLIDAR with T2D. Besides being the current state-of-the-art on such data it is also, among all previous works, the closest to our own. Again, it consists of separate VAD, clustering-based diarization, and a speaker-conditioned E2E joint SD+ASR model. It also has been pre-trained with significantly more supervised data (including $75$k~\cite{kanda2020joint} proprietary data). Instead, our model mostly leverages the WavLM-large self-supervised pre-training.
We include in the comparison also two different clustering-based diarization methods: VBx~\cite{landini2022bayesian} and  overlap-aware spectral clustering~\cite{raj2021multi}. 

Regarding cpWER, we can see that SLIDAR, compared to T2D, in general obtains slightly better figures. We also report, for SLIDAR, figures obtained by using oracle diarization as a prompt (\texttt{<|OD|>} mode, see Sec.~\ref{ssec:multi_task}) for the current window prediction. 
We can see that, if oracle diarization is made available, cpWER can be reduced quite dramatically on SLIDAR. Further work could explore a two-pass decoding approach, using the previously estimated diarization output in place of oracle one.
%Instead T2D is not designed to fully exploit oracle diarization but using oracle \ac{VAD} with T2D brings a minor improvement.% Given this promising results it may be worth exploring in future a two-pass decoding approach where, in the second pass \texttt{<|OD|>} with the previously estimated diarization output is used. 

In general, SLIDAR has worse performance than T2D regarding diarization. %and is even surpassed by VBx (albeit with oracle \ac{VAD}) in the IHM-MIX scenario. 
SLIDAR tends to produce higher false alarms (FA) compared to T2D as it over-estimates utterance boundaries. 
This may be due to the fact that SLIDAR has been trained with manual annotations from a variety of corpora on top of AMI. These may have mismatched segmentation compared to the one used here for evaluation on AMI (which, again, has no DER forgiveness collar). For example, LibriSpeech and CH\-iME-6 manual segmentation is generally less tight. 
%while T2D, being trained with word-level segmentation 
%This is due to the fact it has been not trained with FA-derived AMI segmentation but manual one and, as such, evaluation protocol and training here are mismatched for diarization. 
%In fact, such DER is computed with no collar, an arguable evaluation setup as segmentation is often difficult to define precisely. 
%arguably a rather artificial evaluation setup as such tight segmentation is hardly necessary (and difficult to define precisely) in real-world applications.  

%This promising result,  future works we could explore a two-pass decoding strategy for SLIDAR, where the diarization output of the first pass is used in place of oracle diarization in the second pass.

\begin{table}[H]
\vspace{-0.4cm}
	\centering
	\setlength{\tabcolsep}{2pt} % Default value: 6pt
	\renewcommand{\arraystretch}{1} % Default value: 1
	\small
	\caption{DER (\%) and cpWER (\%) on the AMI corpus IHM-MIX recordings for \texttt{dev} and \texttt{eval} sets. We study the impact of accurate vs. approximate word-level segmentation for truncation (Sec.~\ref{ssec:local_dast}) and of the different multi-task training strategies (Sec.~\ref{ssec:multi_task}).}

	\begin{tabular}{clcccc}
		%    \multicolumn{6}{c}{\normalsize Synthetic Dataset}\\
		\toprule
		\multicolumn{2}{c}{{ Configuration }} & \multicolumn{2}{c}{{ dev }} &  \multicolumn{2}{c}{{ eval }} \\
		\cmidrule{3-4} \cmidrule{5-6} 
		& & DER & cpWER & DER & cpWER \\
		\midrule
            & M2T (FA) & 22.8 & 16.5 & 24.8 & 17.9 \\
		\multirow{4}{*}{} & M2T (approximate) & 26.7 & 16.7 & 28.1 & 18.2 \\
		& + \texttt{<|vanillaASR|>} & 25.8 & 15.1  & 27.3 & 15.9 \\
		& + \texttt{<|prev|>}  & 25.4 & 14.4 & 26.8 & 15.5 \\
            & + \texttt{<|OD|>} (train-only) & 25.2 & 14.2 & 27.1 & 15.6 \\
	    \bottomrule
	\end{tabular}
	
	\label{tab:ablation_ihm}
	\vspace{-0.2cm}
\end{table}
%\begin{figure}[!htbp]
%\centering
%\includegraphics[width=0.4\textwidth]{perf_vs_win.pdf}
%\vspace{-0.4cm}
%\caption{Effect of inference window size on DER (\%) and cpWER (\%) on the AMI IHM-MIX eval set.  SC: speaker confusion (\%) , MS: missed detection (\%) , FA: false alarm (\%).}
%\label{fig:perf_vs_window}
%\end{figure}

In Table~\ref{tab:ablation_ihm}, we report an ablation study performed on AMI IHM-MIX. We quantify: (i) the impact of various multi-task/prompting strategies; (ii) having approximate vs. using \ac{FA}-obtained segmentation for the purpose of utterance truncation (see Sec.~\ref{ssec:local_dast}) and as utterance segmentation ground truth. 
Training with FA-obtained segmentation has little impact on cpWER but has a larger impact on the DER, which decreases considerably. Again, this may be because it reduces the training versus evaluation segmentation annotation mismatch.
%This confirms what has been mentioned before about segmentation annotation mismatch between training and evaluation. %and confirms our heuristic for approximate truncation strategy.
Among the multi-task objectives, \texttt{<|vanillaASR|>} has the most impact, as additional (non multi-speaker annotated, see~\cite{cornell2023chime}) Mixer 6 data can be leveraged.
Adding \texttt{<|OD|>} in training appears to degrade performance negligibly but, as a trade-off, we gain additional model flexibility. 

%as DER is again computed with no collar. 
%We can see that for shorter inference window lengths the performance in terms of mostly recognition (cpWER) degrades. 
%On the other hand, the effect on diarization is more limited. 
%We can thus conclude that there is a trade-off between computational cost and performance.
%In fact, as we are using a self-attention based model (see Sec.~\ref{ssec:model_arc}), using a longer inference window results in a quadratically greater memory cost.

%In particular, for SASSR we report additional results obtained by using oracle diarization prompting also in inference (\texttt{<|OD|>} mode, see Sec.~\ref{ssec:multi_task}) and by using a two-pass decoding strategy enabled by the former. 
%In this two-pass decoding strategy, we derive diarization information in the first pass and then feed it via prompting to the model in place of the oracle. We did not explore further combination e.g. via ROVER~\cite{fiscus1997post}. 
%We can see that 

\section{Conclusions}\label{sec:conclusions}
In this work, we presented SLIDAR, a framework for ``quasi" E2E joint speaker diarization and ASR (SD+ASR) for long-form recordings. This framework employs a sliding window strategy where a fully E2E SD+ASR local model is applied on each window and the global speaker tracking is demanded to a clustering mechanism. 
Our experiments performed on AMI IHM-MIX and SDM mix show promising results. SLIDAR compares favorably with state-of-the-art methods such as Transcribe-to-Diarize, despite using significantly less supervised training data and offering a much more streamlined, almost E2E pipeline. Future research directions include modifying this framework for streamable inference and trying to devise a way for fully E2E long-form DAST. This latter requires to devise a memory-efficient E2E learnable mechanism for long-term speaker tracking. 
%by trying to replace the clustering operation with an E2E learnable mechanism. 

%talk about iterative processing and extension to multiple mics. 
%talk about e2e vs interpretable and composable trend. 
%efficient attention ? not really end2end clustering is not e2e 
%problem with using pre-trained models 
%streamable inference 
%\vfill\pagebreak


\clearpage
\bibliographystyle{IEEEtran}
{\footnotesize
  \setstretch{0.9}
\bibliography{refs.bib}
 }

\end{document}